\documentclass[aps,twocolumn,superscriptaddress,preprintnumbers,showpacs,floatfix,article,prb]{revtex4-1}

\usepackage[latin9]{inputenc}
\usepackage{color}
\usepackage{verbatim}
\usepackage{amsmath}
\usepackage{amssymb}
\usepackage{esint}
\usepackage{esvect}
\usepackage{braket}
\usepackage{amsmath,bm}
\usepackage{url}
\usepackage{lipsum}
\usepackage{lipsum}
\usepackage[dvipsnames]{xcolor}
\usepackage{hyperref}
\usepackage{etoolbox}
\usepackage{lipsum}
\usepackage{bm}
\usepackage{dsfont}
\usepackage{mathtools}

\makeatletter

\@ifundefined{textcolor}{}
{
 \definecolor{BLACK}{gray}{0}
 \definecolor{WHITE}{gray}{1}
 \definecolor{RED}{rgb}{1,0,0}
 \definecolor{GREEN}{rgb}{0,1,0}
 \definecolor{BLUE}{rgb}{0,0,1}
 \definecolor{CYAN}{cmyk}{1,0,0,0}
 \definecolor{MAGENTA}{cmyk}{0,1,0,0}
 \definecolor{YELLOW}{cmyk}{0,0,1,0}
}

\DeclarePairedDelimiterX\inp[2]{\langle}{\rangle}{#1 \delimsize\vert #2}

\DeclarePairedDelimiterX{\Exp}[1]{\langle}{\rangle}{#1}

\usepackage{mathrsfs}

\draft
\makeatother
\begin{document}

\title{Optical rotation in thin chiral/twisted materials and the gyrotropic magnetic effect}
\author{Yan-Qi Wang}
\affiliation{Department of Physics, University of California, Berkeley, CA 94720, USA}
\affiliation{Materials Sciences Division, Lawrence Berkeley National Laboratory, Berkeley CA 94720, USA}
\author{Takahiro Morimoto}
\affiliation{Department of Applied Physics, The University of Tokyo, Hongo, Tokyo, 113-8656, Japan}
\affiliation{JST, PRESTO, Kawaguchi, Saitama, 332-0012, Japan}
\author{Joel E. Moore}
\affiliation{Department of Physics, University of California, Berkeley, CA 94720, USA}
\affiliation{Materials Sciences Division, Lawrence Berkeley National Laboratory, Berkeley CA 94720, USA}

\begin{abstract}
The rotation of the plane of polarization of light passing through a non-magnetic material is known as natural optical activity or optical gyrotropy.  The behavior of this effect in thin chiral conductors is of current interest.  For example, the low frequency limit of gyrotropy in chiral 3D crystals, known as the gyrotropic magnetic effect (GME), is controlled by the orbital magnetic moment of electrons, which has been proposed to be relevant to current-induced switching in twisted bilayer graphene.  We show that the GME is not limited to bulk materials but also appears for quasi-2d systems with minimal structure incorporated in the third direction.  Starting from multi-band Kubo formula, we derive a generic expression for GME current in quasi-2d materials induced by low-frequency light, and provide a Feynman-diagrammatic interpretation. The relations between the 2d finite layered formula and 3d bulk formula are also discussed.
\end{abstract}

\maketitle
\section{Introduction}
The breaking of symmetry in a medium between polarized light with different chiralities is called optical gyrotropy, leading to the rotation of the polarization plane in propagation~\cite{Landau1984}. Faraday rotation and magnetic circular dichroism, for instance, are time-reversal-odd gyrotropic effects and appear at zeroth order in the wave vector of light. At linear order in wave vector, the time-reversal-even part of the optical response is called natural gyrotropy, whose dissipative part leads to natural circular dichroism, while the reactive part gives the optical rotation known as natural optical activity~\cite{Landau1984,Orenstein2013,Zhong2016,Agranovich1984}.

Recently the mechanism of these effects at low frequencies has been of considerable interest.  Beyond merely probing a material's symmetry, the low-frequency limit of natural gyrotropy in chiral 3D metals turns out to probe a very basic property of Bloch electrons, in a loosely similar way to electric polarization and other Berry-phase phenomena.  This limit was named the gyrotropic magnetic effect (GME)~\cite{Zhong2016}, as it includes as a special case one version of the previously discussed chiral magnetic effect in Weyl semimetals~\cite{Son1012,Goswami2013,Chang2015,Ma2015}. It is controlled by the intrinsic orbital magnetic moment of the electrons on the Fermi surface~\cite{Ma2015,Zhong2016,Flicker2018}, which is determined by the Bloch states and is related to but distinct from the more familiar Berry curvature. 
The GME has recently been studied in Weyl semimetals by first-principle calculations~\cite{Goswami2015,Tsirkin2018}.  While there are clearly a variety of nonlinear effects in Weyl semimetals known to be interesting and even approximately quantized~\cite{Son1012,Vazifeh2013,Yamamoto2015,Chen2013,Chang2015,Goswami2013,Wu2017,Fernando2017,Rees2019, Avdoshkin2019}, the GME remains a relatively straightforward probe of chirality as it is a linear response.

The motivation for this paper is to understand how the simplest electromagnetic response to chirality in a time-reversal-invariant system, the GME, is modified in a minimal chiral structure, such as a stack of a few rotated layers, rather than a bulk crystal.  The optical phenomena in thin conductors, with minimal structure incorporated in the third direction, is of great current interest~\cite{Havener2012,Kim2016}. One of the platforms is twisted bilayer graphene~\cite{Santos2007,Bistritzer2011,Santos2012,Cao2018A,Cao2018B,Xu2018,Po2018,Yuan2018,Koshino2018,Isobe}. The electronic structure of this kind of quasi-two-dimensional system is significantly modified by the Moir\'e superlattice, leading to almost flat bands.

The non-trivial Berry phase of the flat bands generates large out-of-plane orbital magnetic moment~\cite{Thonhauser2005,Ceresoli2006,Xiao2005,Shi2007}, which is believed to be relevant to switching in twisted bilayer graphene~\cite{He2019,Sharpe605,Serlin2019}.  The orbital moment's effect on light propagating in the plane of a thin structure is fairly straightforward, but for light passing through the plane, the situation is more complex.  The quasi-2d chiral structure should give rise to optical gyrotropy on symmetry grounds, with some connection to the in-plane orbital magnetic moment, which still remains obscure due to the breaking of translation symmetry along the out-of-plane direction~\cite{Bianco2011,Bianco2013}.

In this article, we start from a standard multiband Kubo formula to derive a generic expression for GME current (or equivalently optical rotation) in quasi-2d materials induced by low-frequency light. We show that the orbital magnetic moment can be expressed in terms of the position operator in the presence of open boundary condition. Similar to recent work on nonlinear optical responses with respect to electric field~\cite{Dan2019}, we provide a diagrammatic interpretation for the Kubo formula results, which in this case should be viewed as responding to magnetic field.  By stacking  quasi-2d layers periodically along the out-of-plane direction, one obtains a thermodynamic limit in which the system is equivalent to a 3d bulk material. The relations between the 2d finite layered formula and 3d bulk formula are also discussed, both analytically and numerically.  Optical rotation is a powerful and widely used probe of chirality of quasi-2d materials, and we hope that our results will extend this technique from simply a probe of symmetry, or the sign of twisting, to a more quantitative probe of electronic chirality.

The paper is structured as follows. In Sec.~\ref{Preliminary}, we briefly introduce the model Hamiltonian, as well as basic properties of the GME coefficient and optical rotation. In Sec.~\ref{Kubo}, we derive a formula for GME coefficient of quasi-2d material, and show the emergence of the position operator in the low-frequency limit. In Sec.~\ref{Diagrammatic}, we give a diagrammatic interpretation of a formula derived from Sec.~\ref{Kubo}. In Sec.~\ref{Convergence}, we use the 2d formula to calculate the GME coefficient for periodic stacked many-layer system, and show its convergence to 3d bulk results in thermodynamic limit. We discuss and summarize the main results in Sec.~\ref{Discussion}, with an eye towards future applications. 

\begin{figure}[!t]
\centering 
\includegraphics[width=1\columnwidth]{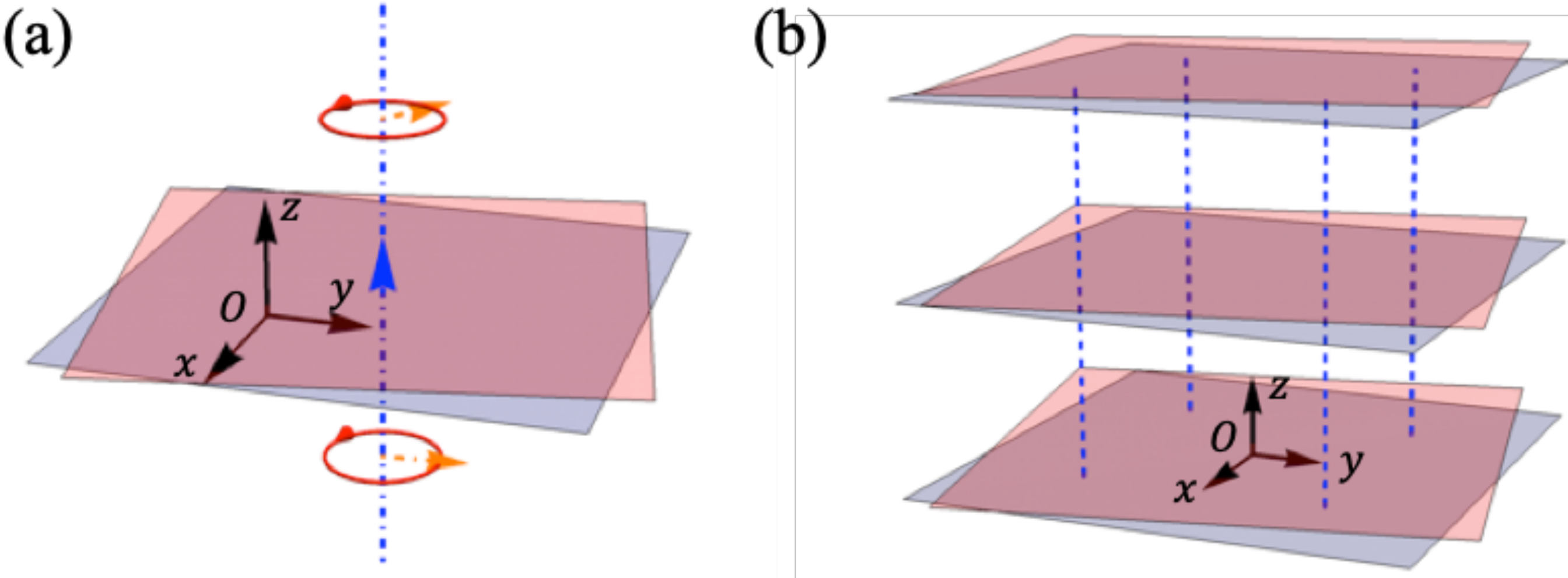}
\caption{\label{OpticalRotation} (a) The rotation of the polarization plane of light passing through a quasi two dimensional chiral material (see the blue and red sheets). The blue dashed arrow perpendicular to the plane denotes the direction of the light. The orange arrows show the rotation of the polarization plane.  (b) Stacking of quasi-2d layers with chiral structure described in (a), the blue dahsed lines illustrate the interlayer couplings.}
\end{figure}

\section{Preliminaries: the Hamiltonian and the GME}\label{Preliminary}
The rotation of the plane of polarization of light passing through a non-magnetic material is known as natural optical activity or optical gyrotropy. Consider the optical gyrotropy for chiral quasi-2d materials on $x-y$ plane. Here the term ``quasi-2d'' means that the system is infinite in the $(x,y)$ plane with well-defined $(k_x,k_y)$, while maintaining finite open boundary conditions along the $\hat z$ direction, which encoded inhomogeneous structure.  Fig.~\ref{OpticalRotation}(a) shows the minimal structure of a single (effectively) chiral layer which consists two sheets (see the blue and red planes). On the other hand, one can stack and couple $N_z$ copies of structure mentioned in Fig.~\ref{OpticalRotation}(a) along the $\hat z$ direction, making a 3d bulk chiral material in the thermodynamic limit, as shown in Fig.~\ref{OpticalRotation}(b).  For such materials, optical rotation at low frequencies has been shown to be related to the orbital magnetic moment of 3d Bloch electrons on the Fermi surface~\cite{Ma2015,Zhong2016}. 

\subsection{The model}\label{Model}
Let us first consider the following Pauli Hamiltonian for a free fermion with spin-orbit coupling~\cite{BLOUNT1962305,Zhong2016}, which will be the starting point for either of the aforementioned cases (single-layer or many-layer slabs):
\begin{equation}\label{2dRealH0}
	{\mathcal H}_0 = \frac{{\bf p}^2}{2m} + V({\bf r}) + \frac{\hbar}{4m^2c^2} ({\bf p} \cdot{} {\bm \sigma} )\times {\bm \nabla} V({\bf r}),
\end{equation}
with $m$ the electron mass, ${\bf p}$ the electron momenta, $c$ the speed of light, ${\bm \sigma} = (\sigma_x, \sigma_y, \sigma_z)$ the Pauli matrices, and $V({\bf r})$ the lattice potential. Here and after we set $c = 1$ for simplicity.  The kinematic momentum associated with ${\mathcal H}_0$ is then defined as:
\begin{equation}\label{2dRealInt}
	\hat {\bm \pi} = \frac{m}{i\hbar}[\hat {\bf r}, {\mathcal H}_0] = \hat {\bf p} + \frac{\hbar}{4m}{\bm \sigma} \times \nabla V({\bf r}),
\end{equation}
which satisfies the commutation relation $[\hat r_i, \hat \pi_j] = [\hat r_i, \hat p_j] = i\hbar \delta_{ij}$. To the leading order of vector potential ${\bf A}({\bf r}, t) = {\bf A}(\omega, {\bf q}) e^{i {\bf q} \cdot{} {\bf r} - i \omega t}$, with ${\bf q} = (q_x,q_y,q_z)$, the coupling with an external electromagnetic field can be treated as a perturbation:
\begin{equation}
	{\mathcal H}_{\text{I}} \approx \frac{e}{2}(\hat {\bf v} \cdot{} {\bf A} ({\bf r}, t) + {\bf A}({\bf r} ,t) \cdot{} \hat {\bf v}) + \frac{g_s e}{2m} [{\bm \nabla} \times {\bf A}({\bf r},t)] \cdot{} {\bf S},
\end{equation}
with ${\bf S} = (\hbar /2) {\bm \sigma}$ being the spin operator. Here the $\hat {\bf v} = \hat {\bm \pi}/m$ is the velocity operator without the external field, whose Fourier transformation is defined as: ${\bm \pi}_{\bf q} = e^{- i {\bf q} \cdot{} {\bf r}} \hat {\bm \pi} e^{+ i {\bf q} \cdot{} {\bf r}} $. Thus the total Hamiltonian reads:
\begin{equation}
	{\mathcal H} = {\mathcal H}_0 + {\mathcal H}_{\text{I}}.
\end{equation}
The velocity operator in the presence of electromagnetic fields can be defined as following: 
\begin{equation}
	\hat {\bf v}^{\text{tot}} = \frac{1}{i\hbar} [{\bf r}, {\mathcal H}] = \frac{\hat {\bm \pi}}{m} + \frac{e{\bf A}({\bf r}, t)}{m}.
\end{equation}
One can transform the Hamiltonian ${\mathcal H}_0$ into Fourier space $H_0(k_x,k_y)$, whose eigenstates $\ket{\Psi_{\bf k}^j} = \ket{\Psi_{k_x,k_y}^j} = {\mathcal V}^{-1/2}e^{ik_x x + ik_yy} \ket{u_{k_x,k_y}^j(z)}$, i.e., the 2d Bloch states satisfy $H_0(k_x,k_y)\ket{\Psi_{k_x,k_y}^j} = E_{k_x,k_y}^j \ket{\Psi_{k_x,k_y}^j}$. The Bloch states are normalized for the entire volume ${\mathcal V}$ of the 3d slab: $\langle \Psi_{\bf k}^m | \Psi_{{\bf k}^\prime}^n \rangle_{\text{All}} = \delta_{mn} \delta_{{\bf k}, {\bf k}^\prime}$. Note that $j$ is the band index which mixes the spin, orbital and sheet/layer structure encoded along $\hat z$ direction.

Now let us assume that in tight-binding limit the Hamiltonian is expanded under maximally localized Wannier functions $\ket{\phi^j_s}$ which can diagonalize the position operator $\hat z \ket{\phi_s^j} = (R_{s} + r_{j}) \ket{\phi_s^j}$.  Here $s$ labels the unit cell and $j$ labels the generalized orbital within the unit cell.  $R_{s}$ denotes the $z$ position for the center of $s$-th unit cell, while $r_{j}$ is the $z$ position for the center of orbital with respect to the center of the unit cell. The Bloch-like basis reads: $\ket{\chi_k^j} = \sum_s e^{ik(R_s + r_j)} \ket{\phi_s^j}$, and the Bloch states can be expanded as: $\ket{\Psi_{\bf k}^n} = {\mathcal V}^{-1/2}\sum_j C_{{\bf k},j}^n \ket{\chi_{\bf k}^j}$.  We note that here we excluded certain topologically nontrivial states such as Chern insulators for which not all states can be localized.

\subsection{The GME coefficient}
 In this section, we briefly introduce some concepts and notations from the gyrotropic magnetic effect~\cite{Zhong2016,Hornreich1968,Malashevich2010}. Note that, to linear order, the total current density induced by a monochromatic light wave reads:
\begin{equation}
	j_i({\bf q}) = \Pi_{ij}({\bf q}) {\bf A}_j({\bf q}),
\end{equation}
and optical gyrotropy is described by the anti-symmetric part of the response tensor $\Pi^A_{ij} = (\Pi_{ij} - \Pi_{ji})/2$ to the order ${\mathcal O}({\bf q})$~\cite{Landau1984,Zhong2016}. Its Taylor expansion to  first order in ${\bf q}$ captures the natural gyrotropy:
\begin{equation}
	\Pi^A_{ij}({\bf q}) = \Pi^A_{ij}(0) + \Pi^A_{ijl}q_l + \cdots.
\end{equation}
The time-reversal-even part $\Pi^A_{ijl}$ (GME tensor) with nine independent component is antisymmetric under $ i \leftrightarrow j$ and can be written using the GME coefficient $\alpha_{ij}^{\text{GME}}$, where the latter is a rank two tensor:~\cite{Zhong2016,Hornreich1968,Malashevich2010}
\begin{subequations}
\begin{align}
		\Pi^A_{ijl} &= i \epsilon_{ilp} \alpha_{jp}^{\text{GME}} - i\epsilon_{jlp}\alpha_{ip}^{\text{GME}} \\
		\alpha_{ij}^{\text{GME}} &= \frac{1}{4i} \epsilon_{jlp} (\Pi^A_{lpi} - 2\Pi^A_{ilp}). \label{tensor}
\end{align}
\end{subequations}
In the low-frequency limit $\hbar \omega \ll E_{\text{gap}}$ ($E_\text{gap}$ stands for the band gap) where only the intra-band absorption can occur, we further have:
\begin{equation}
	j_i = \alpha_{ij}^{\text{GME}} B_j.
\end{equation}

For a metal with cubic symmetry or higher, one can derive $\Pi^A_{ijl} = -2i \bar\alpha^{\text{GME}}\epsilon_{ijl}$ and $\alpha_{ij}^{\text{GME}} = \bar \alpha^{\text{GME}} \delta_{ij}$. With this the rotatory power $\rho$ can be expressed in terms of ${\bm \alpha}^{\text{GME}}$:~\cite{Agranovich1984,Zhong2016}

\begin{equation}
	\rho = - (1/ \epsilon_0 ) \text{Re} \bar \alpha^{\text{GME}}.
\end{equation}
In general the rotatory power has the unit of rad/unit length. According to Eq.~(\ref{tensor}), $\alpha_{xx}^{\text{GME}}$ can be expanded by $\Pi_{ijl}^A$:
\begin{equation}\label{alphapi}
	\alpha_{xx,2d}^{\text{GME}} =  - \frac{i}{2} \Pi_{yzx}^A + \frac{i}{2}\Pi_{xyz}^A - \frac{i}{2}\Pi^A_{xzy},
\end{equation}
where those GME tensors can be derived from the standard perturbation theory.  One thing we wish to recall for clarity is that the quasi-2d material here still lives in three dimensions, and the only fundamental difference is the breaking of translation symmetry along the out-of-plane direction. The $\alpha_{xx,2d}^{\text{GME}}$ we defined here has the same units compared with the $ \alpha_{xx,3d}^{\text{GME}}$ for 3d bulk material, aside from the unit length along the third direction is switched to the thickness of the slab. For example, the rotatory power for a 3d bulk material by stacking infinite many structures plotted in Fig.~\ref{OpticalRotation}(a) has the units of rad/$a$, with $a$ is the lattice constant along $\hat e_z$ direction. On the other hand, the rotatory power for a quasi-2d slab with 3 unit layers shown in Fig.~\ref{OpticalRotation}(b) has more naturally the units of rad/$3a$ since there is no true unit cell or periodicity along $\hat e_z$. The main goal for this paper is to calculate $\alpha_{xx,2d}^{\text{GME}}$ (or $\alpha_{yy,2d}^{\text{GME}}$) for a quasi-2d system, which characterizes the rotation of the polarization plane of light perpendicular to the quasi-2d slab.

\section{Generic Kubo formula}\label{Kubo}
Based on standard perturbation theory, we first derive the generic 2d formula for anti-symmetric conductance tensor $\Pi_{\alpha \beta}^A$ as a function of wave vector of light in Sec.~\ref{2d}. Then, in Sec.~\ref{cxyz}  and Sec.~\ref{cyzx}, at low frequency limit we evaluate the GME tensor $\Pi_{xyz}^A$ and $\Pi_{yzx}^A/\Pi_{xzy}^A$ in terms of position operator $\hat z$. We derive the $\alpha_{xx}^{\text{GME}}$ (Eq.~(\ref{alphaxx})) for 2d material at the end of the section, which is the main general result of this paper.

\subsection{Response tensor for 2d material}\label{2d}

We start this section from standard perturbation theory. We first derive the net-current, then we treat the electro-magnetic field as the perturbation, and evaluate the perturbative matrix element restricted by photon-momenta transfer. Finally we arrive at the main result of this section, which is the rank-2 anti-symmetric conductance tensor $\Pi_{\alpha \beta}^A$, see in Eq.~(\ref{pialphabeta}).

\subsubsection{The net-current} 
Based on standard perturbation theory~\cite{Wooten,Dressel2002,Harrison1980,Melrose1991}, the 2d current density induced by the monochromatic light ${\bf A}({\bf r}, t) = {\bf A}(\omega, {\bf q}) e^{i{\bf q} \cdot{} {\bf r} - i\omega t}$ reads:
\begin{equation}\label{TotalCurrent}
\begin{aligned}
	\hat J({\bf q}) &= -\frac{e}{2} \text{Tr}\big{\{} [\hat {\bf v}^{\text{tot}}e^{- i {\bf q \cdot{} {\bf r}}}  +e^{-i {\bf q} \cdot{} {\bf r}} \hat {\bf v}^{\text{tot}} ] {\mathcal N} \big{\}} \\
	&\approx -\frac{e^2}{m} \text{Tr} \big{\{} e^{- i {\bf q} \cdot{} {\bf r}} {\mathcal N}_0 {\bf A} ({\bf r}, t)\big{\}} \\
	&~~~~ - e~\text{Tr} \big{\{}  \big{[} \hat {\bf v} e^{-i {\bf q} \cdot{} {\bf r}}  +  e^{- i {\bf q} \cdot{} {\bf r}} \hat {\bf v}   \big{]} \delta {\mathcal N} \big{\}}/2,
\end{aligned}
\end{equation}
where the trace and integral is conducted in ``All'' space ${\mathcal V}$ under the quasi-2d Bloch states $\ket{\Psi_{\bf k}^j}$ defined in Sec.~\ref{Model}. Here the particle density ${\mathcal N}$ has been decomposed into the unperturbed density ${\mathcal N}_0$ and the density fluctuation induced by interaction: ${\mathcal N} = {\mathcal N}_0 + \delta {\mathcal N}$.

One can also decompose the total current Eq.~(\ref{TotalCurrent}) as $\hat {J}({\bf q}) = \hat J^1({\bf q}) + \hat J^2({\bf q})$. The first term $ \hat J^1({\bf q})$ is the so called dia-magnetic term:
\begin{equation}\label{Diamagnetic}
	\hat J^1({\bf q}) = - \frac{e^2}{mc} \text{Tr}\big{\{} e^{- i {\bf q} \cdot{} \bf r} {\mathcal N}_0 {\bf A}({\bf r}, t) \big{\}} = - \frac{e^2}{mc} \sum_{j} f(E_j){\bf A},
\end{equation}
where $f(E_j) = 1/(1 + e^{(E_j - \mu)/k_BT})$ is the Fermi distribution function for the system with chemical potential $\mu$. Hereafter, we simply write ${\bf A}$ for ${\bf A}(\omega, {\bf q})$.  

Now we want to evaluate the second term in Eq.~(\ref{TotalCurrent}):
\begin{equation}
	\hat J^2({\bf q})= - \frac{e}{2}\text{Tr}\big{\{} [\hat {\bf v} e^{-i {\bf q} \cdot{} {\bf r}}  +  e^{- i {\bf q} \cdot{} {\bf r}} \hat {\bf v} ]  \delta {\mathcal N}\big{\}}.
\end{equation}
We insert a complete set ${\mathds 1} = \sum_{{\bf k}^\prime,j^\prime}\ket{\Psi_{{\bf k}^\prime}^{j^\prime}} \bra{\Psi_{{\bf k}^\prime}^{j^\prime}}$ inside: $\hat J^2({\bf q}) = - \frac{e}{2} \sum_{j,j^\prime}\sum_{{\bf k}^\prime,{\bf k}}\bra{\Psi_{{\bf k}}^j} [ \hat {\bf v} e^{-i {\bf q} \cdot{} {\bf r}}  +  e^{- i {\bf q} \cdot{} {\bf r}} \hat {\bf v} ]   \ket{\Psi_{{\bf k}^\prime}^{j^\prime}} \bra{\Psi_{{\bf k}^\prime}^{j^\prime}} \delta {\mathcal N} \ket{\Psi_{{\bf k}}^j}$. We will consider the $\bra{\Psi_{{\bf k}}^j} [ \hat {\bf v} e^{-i {\bf q} \cdot{} {\bf r}}  +  e^{- i {\bf q} \cdot{} {\bf r}} \hat {\bf v} ]   \ket{\Psi_{{\bf k}^\prime}^{j^\prime}}$ in Sec.~\ref{PertMatrix}.

The matrix element $\bra{\Psi_{{\bf k}^\prime}^{j^\prime}} \delta {\mathcal N} \ket{\Psi^j_{{\bf k}}}$ can be derived from Schrodinger equation under adiabatic approximation~\cite{Zhong2016,Dressel2002,Allen2006}, with $\eta = 1/\tau$ interpreted as the scattering rate:
\begin{equation}\label{Expectation}
	\bra{\Psi_{{\bf k}^\prime}^{j^\prime}} \delta {\mathcal N} \ket{\Psi_{{\bf k}}^j} =  \frac{f^0(E_{{\bf k}^\prime}^{j^\prime}) - f^0(E_{{\bf k}}^{j})}{E_{{\bf k}^\prime}^{j^\prime} - E_{{\bf k}}^j - \hbar \omega - i \hbar \eta} \bra{\Psi_{{\bf k}^\prime}^{j^\prime}} {\mathcal H}_{\text{I}} \ket{\Psi_{{\bf k}}^j},
\end{equation}
Following some well known tricks in the low-frequency limit\cite{Allen2006,Zhong2016}, we find that the prefactor $ {[f^0(E_{{\bf k}^\prime}^{j^\prime}) - f^0(E_{{\bf k}}^{j})]}/{[E_{{\bf k}^\prime}^{j^\prime} - E_{{\bf k}}^j - \hbar \omega - i \hbar \eta]}  $ can be divided into two parts:
\begin{equation}\label{decompose}
 \frac{\hbar \omega[f^0(E_{{\bf k}}^j) - f^0(E_{{\bf k}^\prime}^{j^\prime})]}{E_{{\bf k}}^j - E_{{\bf k}^\prime}^{j^\prime} + i\hbar \eta} \bigg{(} \frac{1}{\hbar \omega} -\frac{1}{\hbar \omega + E_{{\bf k}}^j - E_{{\bf k}^\prime}^{j^\prime}} \bigg{)}.
\end{equation}
The first term in the Eq.~(\ref{decompose}) can be viewed as: ${\mathcal P} \big{[} (f^0(E_{{\bf k}}^j) - f^0(E_{{\bf k}^\prime}^{j^\prime}))/(E_{{\bf k}}^j - E_{{\bf k}^\prime}^{j^\prime}) \big{]} + i \pi \big{[} f^0(E_{{\bf k}}^j) - f^0(E_{{\bf k}^\prime}^{j^\prime})\big{]}\delta(E_{{\bf k}}^j - E_{{\bf k}^\prime}^{j^\prime})$, with the ${\mathcal P}$ stands for Cauchy principle value. The term related to principle value cancelled with the diamagnetic term $\hat J^1({\bf q})$ (Eq.~(\ref{Diamagnetic})). The second term vanishes, since when $\delta(E_{{\bf k}}^j - E_{{\bf k}^\prime}^{j^\prime}) = 1$, we have $f^0(E_{{\bf k}}^j) - f^0(E_{{\bf k}^\prime}^{j^\prime}) =0$~\cite{Allen2006,Zhong2016}. Thus, by combining $\hat J^1({\bf q})$ and $\hat J^2({\bf q})$, only the second term of Eq.~(\ref{decompose}) contributes to the net current:
\begin{equation}\label{ParaCurrent}
\begin{aligned}
	{\hat J({\bf q})} = &- \frac{e}{2} \sum_{j,j^\prime}\sum_{{\bf k}^\prime,{\bf k}}\bra{\Psi_{{\bf k}}^j} [ \hat {\bf v} e^{-i {\bf q} \cdot{} {\bf r}}  +  e^{- i {\bf q} \cdot{} {\bf r}} \hat {\bf v} ]   \ket{\Psi_{{\bf k}^\prime}^{j^\prime}} \langle {\delta \mathcal N} \rangle_{\bf k^\prime {\bf k}}^{j^\prime j}.
\end{aligned}
\end{equation}
Here we define: $\langle \delta {\mathcal N} \rangle_{{\bf k}^\prime {\bf k}}^{j^\prime j}=F^{j j^\prime}_{{\bf k},{\bf k}^\prime}\bra{\Psi_{{\bf k}^\prime}^{j^\prime}} {\mathcal H}_{\text{I}} \ket{\Psi_{\bf k}^j}$, with 
\begin{equation}
	F^{j j^\prime}_{{\bf k} {\bf k}^\prime} = - \frac{f^0(E_{\bf k}^j) - f^0(E^{j^\prime}_{{\bf k}^\prime})}{E_{\bf k}^j - E_{{\bf k}^\prime}^{j^\prime} + i \hbar \eta} \frac{\hbar \omega}{\hbar \omega + E_{\bf k}^j - E_{{\bf k}^\prime}^{j^\prime} + i\hbar \eta}.
\end{equation}

\subsubsection{Perturbative matrix element and momentum transfer}\label{PertMatrix}
We would like to evaluate the matrix element of ${\mathcal H}_{\text{I}}$ between two Bloch states. For the monochromatic light with wave vector ${\bf q} = (\tilde {\bf q}, q_z ) = (q_x, q_y, q_z)$, and vector potential ${\bf A} = (A_x, A_y, A_z)$, we have the coupling with light as:
\begin{equation}\label{emfield}
\begin{aligned}
	{\mathcal H}_{\text{I}} &= \frac{e}{2} \sum_i \big{[} v_i e^{i(q_xx + q_yy+ q_zz)}  + e^{i(q_xx+ q_yy + q_zz)} v_i\big{]} A_i.
\end{aligned}
\end{equation}
Note that, in a quasi-2d layered system (a slab in $\hat z$ direction), for given cell periodic operator $\hat {\mathcal O}_{\bf k} = e^{-i {\bf k} \cdot{} {\bf r}} \hat {\mathcal O} e^{+ i {\bf k} \cdot{} {\bf r}}$ (say velocity operator $\hat {\bf v}_{{\bf k}}$) we have: 
\begin{equation}\label{AllCellRelation}
	\bra{\Psi_{{\bf k}^\prime}^{j^\prime}} \hat {\mathcal O} e^{i \tilde {\bf q} \cdot{} {\bf r}} \ket{\Psi_{{\bf k}}^{j}}_{\text{All}} = \delta_{k_x,k_x^\prime- q_x}\delta_{k_y,k_y^\prime - q_y} \bra{u_{{\bf k}^\prime}^{j^\prime}} \hat{\mathcal O}_{{\bf k}^\prime}  \ket{u_{{\bf k}}^{j}}_{\text{Cell}}.
\end{equation}
Here ``All'' stands for the entire space ${\mathcal V}$ where the Bloch states is defined, while the ``Cell'' stands for the volume of a quasi-2d unit cell. By applying this relation, i.e., take $\hat {\mathcal O} = \hat {\bf v} e^{iq_zz}$ which is cell periodic in $(x,y)$ plane, then we shall see: $\bra{\Psi_{{\bf k}^\prime}^{j^\prime}} {\mathcal H}_{\text{I}} \ket{\Psi_{{\bf k}}^j}_{\text{All}} = \delta_{{\bf k}, {\bf k}^\prime -\tilde {\bf q}}\bra{u_{{\bf k} +\tilde {\bf q}}^{j^\prime}}{\mathcal H}_{\text{I},{\bf k}+ \tilde {\bf q}/2} \ket{u_{\bf k}^j}_{\text{Cell}}$, with ${\mathcal H}_{\text{I},{\bf k} + \tilde {\bf q}/2,\beta}$ stands for the 2d Fourier transformation of the term associated with $A_\beta$ in Eq.~(\ref{emfield}). Here ${\mathcal H}_{\text{I},{\bf k} + \tilde {\bf q}/2,\beta}$ stands for the 2d Fourier transformation of the term associated with $A_\beta$ in Eq.~(\ref{emfield}). Similar tricks also apply for the matrix element ahead of $\langle {\delta \mathcal N} \rangle_{\bf k^\prime {\bf k}}^{j^\prime j}$ in Eq.~(\ref{ParaCurrent}): $\bra{\Psi_{{\bf k}}^j} [ \hat {\bf v} e^{-i {\bf q} \cdot{} {\bf r}}  +  e^{- i {\bf q} \cdot{} {\bf r}} \hat {\bf v} ]   \ket{\Psi_{{\bf k}^\prime}^{j^\prime}}_{\text{All}}= \bra{u_{\bf k}^j} v_{{\bf k} + \tilde {\bf q}/2,\alpha} e^{-iq_z r_z} +e^{-iq_z r_z} v_{{\bf k} + \tilde {\bf q}/2,\alpha}   \ket{u_{{\bf k} + \tilde {\bf q}}^{j^\prime}}_{\text{Cell}}$. Now we have successfully transformed the full space integral into the cell integral, and illustrated the momentum shift restriction for a quasi-2d Bloch electron's scattering with light.

\subsubsection{The rank-2 GME tensor $\Pi_{\alpha \beta}^A$}
Combined with results in Sec.~\ref{PertMatrix}, one can subtract the conductance tensor $j_\alpha = \Pi_{\alpha \beta} A_\beta$ from Eq.~\ref{ParaCurrent}. The GME tensor is related to its anti-symmetric part $\Pi_{\alpha\beta}^A = (\Pi_{\alpha \beta}  -\Pi_{\beta \alpha})/2$:
%\begin{widetext}	
\begin{align}\label{pialphabeta}
	\Pi_{\alpha\beta}^A &=  ie \sum_{j,j^\prime,{\bf k}}F^{j j^\prime}_{{\bf k} {\bf k + \tilde {\bf q}}} {\mathcal M}_{\alpha \beta}^A, \\
	{\mathcal M}_{\alpha \beta}^A &=  \text{Im} \bra{u_{\bf k}^j} v_{{\bf k} + \tilde {\bf q}/2,\alpha} e^{-iq_z r_z} +e^{-iq_z r_z} v_{{\bf k} + \tilde {\bf q}/2,\alpha}   \ket{u_{{\bf k} + \tilde {\bf q}}^{j^\prime}}
	\nonumber \\
	&\quad\quad \times \bra{u^{j^\prime}_{{\bf k} + \tilde {\bf q}}} {\mathcal H}_{\text{I},{\bf k} +\tilde {\bf q}/2,\beta} \ket{u^j_{\bf k}} / 2.
\end{align}	
%\end{widetext}
Hereafter we drop the subscript ``Cell'' (``All'') if the operator is evaluated under cell periodic part of Bloch wave function (full Bloch wave function).

Note that $F_{j^\prime j}$ does not explicitly depend on $q_z$. One can send $\tilde {\bf q} \rightarrow 0$. In this case ${\bf k}^\prime \rightarrow {\bf k}$. We have assumed that the frequency is so low, such that $\forall {\bf k} = (k_x, k_y),j\neq j^\prime$, $|E_{\bf k}^j - E_{\bf k}^{j^\prime}| \gg \hbar \omega$, thus $F_{j^\prime j}= 0$. For the $ j = j^\prime$, we shall see:
\begin{equation}
\begin{aligned}
	F_{{\bf k} {\bf k + \tilde {\bf q}}}^{j j^\prime} &= - \lim_{\tilde {\bf q} \rightarrow 0} \frac{f^0(E^j_{\bf k}) - f^0(E^j_{{\bf k} - \tilde {\bf q}})}{E_{\bf k}^j - E_{{\bf k} -\tilde {\bf q}}^j} \frac{\hbar \omega}{\hbar \omega + i \hbar \eta} \\
	&= - \frac{\partial f^0(E_{\bf k}^j)}{\partial E_{\bf k}^j} \frac{\hbar \omega}{\hbar \omega + i \hbar \eta} =   \frac{\partial f^0(E_{\bf k}^j)}{\partial E_{\bf k}^j}  \frac{i\omega \tau}{1 - i\omega \tau},
\end{aligned}
\end{equation}
where we have interpreted $\eta = 1/\tau$ as a scattering rate $1/\tau$~\cite{Zhong2016,Allen2006}. In this case $\Pi^A_{\alpha \beta}$ reads:
\begin{equation}\label{pixy}
	\Pi_{\alpha \beta}^A =   \frac{e^2\omega \tau}{1 - i\omega \tau} \sum_{j} \sum_{{\bf k}}  \frac{\partial f^0(E_{\bf k}^j)}{\partial E_{\bf k}^j} {\mathcal M}_{\alpha \beta}^A,
\end{equation}
with  ${\mathcal M}_{\alpha \beta}^A$ given in Eq.~(\ref{pialphabeta}).

\subsection{The GME tensor ${\Pi}_{xyz}^A$}\label{cxyz}
To get ${\Pi}_{xyz}^A$, let us assume that we have light which is not strictly perpendicular to the $(x,y)$ plane. Instead, assuming that ${\bf q} =(q_x, 0, q_z)$, i.e., $\tilde {\bf q} = (q_x, 0)$ with $|\tilde {\bf q}| \ll |q_z|$. We can approximately view ${\bf A} = (0, A_y, 0)$. In this case we shall have ${\bf q} \cdot{} {\bf r} = q_x x + q_zz$. The coupling with light reads: ${\mathcal H}_{\text{I}} = \frac{e}{2c} \big{[} v_y e^{i(q_xx + q_zz)}  + e^{i(q_xx+q_zz)} v_y\big{]} A_y$. Substituting back to Eq.~(\ref{pialphabeta}) we have: ${\mathcal M}_{xy}^A(q_z) = \text{Im} \big{[}\bra{u_{{\bf k}}^j}  \hat v_{{\bf k} + \tilde {\bf q}/2,x} e^{-i q_z z}  \ket{u_{{\bf k} + \tilde {\bf q}}^{j}} \bra{u_{{\bf k} + \tilde {\bf q}}^{j}} \hat v_{{\bf k} + \tilde {\bf q}/2,y}  e^{+ i q_zz}  \ket{u_{\bf k}^j}\big{]}$. We first take the low frequency limit, and then send $\tilde {\bf q} \rightarrow 0$, such that $F_{j^\prime j} = F_{jj} \delta_{j^\prime j} = \delta_{j^\prime j}[\partial  f^0(E_{\bf k}^j)/\partial E_{\bf k}^j] [i\omega \tau/(1 - i \omega \tau)]$. With above we have the conductance tensor as:
\begin{subequations}
\begin{align}
&\Pi_{xy}^A =   \frac{e^2\omega \tau}{1 - i\omega \tau} \sum_{j} \sum_{{\bf k}}  \frac{\partial f^0(E_{\bf k}^j)}{\partial E_{\bf k}^j}{\mathcal M}_{xy}^A(q_z), \label{xyA} \\
&{\mathcal M}_{xy}^A(q_z) = \text{Im} \bra{u_{{\bf k}}^j}  \hat v_{{\bf k},x} e^{-i q_z z}  \ket{u_{{\bf k} }^{j}} \bra{u_{{\bf k} }^{j}} \hat v_{{\bf k},y}  e^{+ i q_zz}  \ket{u_{\bf k}^j}. \label{mxyqz}
\end{align}	
\end{subequations}
Here $\hat v_{{\bf k},x} = \hbar^{-1}\partial_{k_x} H_0(k_x,k_y)$, and $\hat v_{{\bf k},y} = \hbar^{-1} \partial_{k_y} H_0(k_x,k_y)$.  Note that, in Eq.~(\ref{pixy}), we have used the fact that $\hat v_{{\bf k},x}$ and $\hat v_{{\bf k},y}$ does not explicitly contain $z$, such that they commute with $e^{\pm iq_zz}$. We would like to evaluate Eq.~(\ref{pixy}) to the leading order of $q_z$. One can carry out the Taylor expansion as follows: $e^{\pm iq_zz} \approx 1 \pm iq_zz$ in small $q_z$ limit, drop the real part $\bra{u^j_{\bf k}}\hat v_{{\bf k},x} \ket{u^j_{\bf k}} \bra{u^j_{\bf k}} \hat v_{{\bf k},y} \ket{u_{\bf k}^j}$, and insert a complete set between $\hat v_{{\bf k},y}$ and $\hat z$, such that we have
\begin{equation}\label{pixyz}
\begin{aligned}
	{\mathcal M}_{xyz}^A &\approx \sum_{j_1, j_1 \neq j} \text{Re}\big{[} \bra{u_{\bf k}^j} \hat v_{{\bf k},x} \ket{u_{\bf k}^j}\bra{u_{\bf k}^j}\hat v_{{\bf k},y} \ket{u_{\bf k}^{j_1}} \bra{u_{\bf k}^{j_1}} \hat z \ket{u_{\bf k}^j} \\
	&~~~~~~~~~~~~~~-\bra{u_{\bf k}^j} \hat v_{{\bf k},x} \ket{u_{\bf k}^{j_1}} \bra{u_{\bf k}^{j_1}} \hat  z\ket{u^j_{\bf k}} \bra{u^j_{\bf k}} \hat v_{{\bf k},y} \ket{u_{\bf k}^j}  \big{]},
\end{aligned}
\end{equation}
with the $j_1 = j$ term cancel out.

\subsection{The GME tensor $\Pi_{y z x}^A$ and $\Pi_{x z y}^A$.}\label{cyzx}
The calculation of $\Pi_{\alpha z \beta}^A$ [with $(\alpha, \beta) = ( x,y)$ or $(y, x)$] are slightly different from $\Pi_{xyz}^A$ due to the lack of periodicity in $z$ direction. Let us assume the light with wave vector ${\bf q}$, while $\tilde {\bf q} = q_\beta  \hat e_\beta$, $q_z = 0$, and ${\bf A} = (0, 0, A_z)$. Accordingly, the coupling with light reads: ${\mathcal H}_{\text{I}} = \frac{e}{2c}[v_z e^{iq_\beta v_\beta} + e^{iq_\beta v_\beta} v_z]A_z$. Similar to the calculation in $\Pi_{xyz}^A$, by using Eq.~(\ref{pixy}), in the low frequency limit we have:
\begin{equation}
\begin{aligned}\label{azb}
	{\mathcal M}^A_{\alpha z}(q_\beta) &= \text{Im}\big{[} \bra{u^j_{\bf k}} \hat v_{{\bf k} + q_\beta \hat e_\beta/2,\alpha} \ket{u_{{\bf k}^\prime}^{j}} \bra{u_{{\bf k}^\prime}^{j}} \hat v_z \ket{u_{\bf k}^j} \big{]},
\end{aligned}
\end{equation}
with ${\bf k}^\prime = {\bf k} + q_\beta \hat e_\beta$. Here $\hat v_{{\bf k},\alpha} = \hbar^{-1}\partial_{k_\alpha}H_0(k_x,k_y)$ and $\hat {\bf v}_{{\bf k} \pm {\bf q}/2} = \hat {\bf v}_{\bf k} \pm {\bf q}/2m$~\cite{BLOUNT1962305} for $\alpha = x,y$. However, $\hat v_z$ can not be written in this form due to the breaking of translation symmetry along $\hat e_z$ direction, one can treat $\hat v_z$ as in real space since it commutes with $e^{i {\bf k} \cdot{} {\bf r}}$ and $e^{iq_\beta v_\beta}$. 
With above we can expand $\ket{u_{\bf k}^j}$ and $\bra{u_{\bf k}^j}$ to the leading order of $q_\beta$ ($\ket{u_{{\bf k} + q\beta \hat e_\beta }^j} \approx  \ket{u_{\bf k}^j} + \ket{\partial_{k_\beta}u_{\bf k}^j}q_\beta$ and $\bra{u_{{\bf k} + q\beta \hat e_\beta }^j} \approx  \bra{u_{\bf k}^j} + \bra{\partial_{k_\beta}u_{\bf k}^j}q_\beta$), and substitute back into Eq.~(\ref{azb}), we arrive:
\begin{equation}
\begin{aligned}
	{\mathcal M}^A_{\alpha z}(q_\beta) &= \text{Im} \big{[} \bra{ u_{\bf k}^{j}} \hat v_{{\bf k}, \alpha} \ket{ \partial_{k_\beta} u_{\bf k}^{j} }  \bra{u_{\bf k}^{j}} \hat v_z \ket{u^j_{\bf k}} ] q_\beta \\
	&+ \text{Im} \big{[} \bra{u_{\bf k}^j} \hat v_{{\bf k},\alpha} \ket{u_{\bf k}^{j}} \bra{\partial_{k_\beta} u_{\bf k}^{j}} \hat v_z \ket{u_{\bf k}^j}  \big{]}q_\beta + {\mathcal O}(q_\beta^2) ,
\end{aligned}
\end{equation}	
where we have dropped the real part $\bra{u_{\bf k}^j} \hat v_{{\bf k},\alpha} \ket{u_{\bf k}^{j}} \bra{u_{\bf k}^{j}} \hat v_z \ket{u_{\bf k}^j}$. By inserting a complete set inside we have:
\begin{equation}\label{simazb}
\begin{aligned}
	{\mathcal M}_{\alpha z \beta}^A &= \sum_{j_1, j_1 \neq j} \text{Im}\big{[} \bra{ u_{\bf k}^{j} } \hat v_{{\bf k}, \alpha} \ket{u_{\bf k}^{j_1}} \langle u_{\bf k}^{j_1}|{ \partial_{k_\beta} u_{\bf k}^{j}} \rangle  \bra{u_{\bf k}^{j}} \hat v_z \ket{u^j_{\bf k}} \\
	& ~~~~~~~~~~~~~~ - \bra{u_{\bf k}^j} \hat v_{{\bf k},\alpha} \ket{u_{\bf k}^{j}} \langle{ u_{\bf k}^{j}} | \partial_{k_\beta} u_{\bf k}^{j_1} \rangle \bra{u_{\bf k}^{j_1}} \hat v_z \ket{u_{\bf k}^j} \big{]},
\end{aligned}
\end{equation}
where we have used the fact $\langle u_{\bf k}^{j_1}|{ \partial_{k_\beta} u_{\bf k}^{j}} \rangle  = - \langle { \partial_{k_\beta} u_{\bf k}^{j_1}| u_{\bf k}^{j}} \rangle $ such that the $j_1 = j$ terms cancel out.

 Note that, in the presence of open boundary condition or for an infinite system, we have the following relation~\cite{BLOUNT1962305,Yafet1957,Gu2013} stands for $j \neq l$ ($\hat {\bf p} = im[H, {\bf r}]/\hbar$):
\begin{equation}\label{blount}
	\begin{aligned}
		\langle \hat v_z \rangle_{{\bf k}}^{jl} &= \bra{u_{\bf k}^j} \hat v_z \ket{u_{\bf k}^l} = i \Omega_{ {\bf k}}^{jl} \bra{u_{\bf k}^j} \hat z \ket{u^l_{\bf k}},  \\
		\langle u_{\bf k}^j | \partial_{k_\beta} u_{\bf k}^l \rangle &= - \frac{1}{\Omega^{jl}_{{\bf k}}}\langle \hat v_{{\bf k}, \beta} \rangle_{{\bf k}}^{jl} = - \frac{1}{\Omega^{jl}_{ {\bf k}}} \bra{u_{\bf k}^j} \hat v_{{\bf k}, \beta} \ket{u_{\bf k}^l}.\,
	\end{aligned}
\end{equation}
with $\Omega_{{\bf k}}^{jl} = \hbar^{-1}(E_{\bf k}^j  - E_{\bf k}^l)$ and the position operator defined trivially as in Sec.~\ref{Model}. For a finite system with periodic boundary conditions, an additional correction term should be taken into consideration or we need to use the quantum position operator~\cite{Resta1998,Yu2011,Gu2013}, but this is not a case that we consider in this paper. Substituting Eq.~(\ref{blount}) back to Eq.~(\ref{simazb}) we have:
\begin{equation}\label{piazb}
\begin{aligned}
	{\mathcal M}_{\alpha z \beta}^A &= \sum_{j_1, j_1 \neq j} \text{Im} \big{[} -i \langle \hat v_{{\bf k},\alpha} \rangle_{\bf k}^{jj} \langle \hat v_{{\bf k},\beta}\rangle^{jj_1}_{\bf k}  \langle \hat z \rangle_{\bf k}^{j_1j}   \\
	&~~~~~~~~~~~~~~~~~  - \frac{1}{\Omega_{\bf k}^{j_1j}} \langle \hat v_{{\bf k},\alpha} \rangle^{jj_1}_{\bf k} \langle \hat v_{{\bf k},\beta} \rangle_{\bf k}^{j_1j}\langle \hat v_z \rangle_{\bf k}^{jj}    \big{]}.
\end{aligned}
\end{equation} 

Combining Eq.~(\ref{pixyz}), Eq.~(\ref{piazb}) and Eq.~(\ref{alphapi}), we arrive at the GME coefficient:
\begin{widetext}
\begin{equation}\label{alphaxx}
\begin{aligned}
	\alpha_{xx,2d}^{\text{GME}} &= \frac{i \omega \tau}{1 - i\omega \tau} \sum_{\bf k} \sum_j \frac{\partial f^0(E_{\bf k}^j)}{\partial E_{\bf k}^j}  \langle \hat v_{{\bf k},x}\rangle^{jj}_{\bf k} \sum_{j_1, j_1\neq j} \frac{e^2}{2}\text{Re}\big{[} \langle \hat v_{{\bf k},y} \rangle_{\bf k}^{jj_1} \langle \hat z \rangle^{j_1j}_{\bf k} + \langle \hat z \rangle_{\bf k}^{jj_1}\langle \hat v_{{\bf k}, y} \rangle_{\bf k}^{j_1j} \big{]} \\
	&= \frac{i \omega \tau}{1 - i\omega \tau}\frac{e^2}{2} \sum_{\bf k} \sum_j \frac{\partial f^0(E_{\bf k}^j)}{\partial E_{\bf k}^j}  \langle \hat v_{{\bf k},x}\rangle^{jj}_{\bf k}  \text{Re}\big{[} \langle \hat v_{{\bf k},y} \hat z +  \hat z \hat v_{{\bf k}, y} \rangle_{\bf k}^{jj}  - 2\langle \hat z \rangle_{\bf k}^{jj} \langle \hat v_{{\bf k},y} \rangle_{\bf k}^{jj} \big{]}
\end{aligned}
\end{equation}
\end{widetext}
which is the main result of this paper. Note that, due to the orthogonality of eigenstates, the final results do not depend on the chosen zero point of the $\hat z$ coordinate. One can shift the origin of $\hat z$ coordinates by $z_0$, such that under new coordinate we have:  $\bra{u_{\bf k}^{i}}\hat z + z_0{\mathds I}_{4N_z \times 4N_z} \ket{u_{\bf k}^j} = \bra{u_{\bf k}^i}\hat z \ket{u_{\bf k}^j} + z_0\delta_{ij} = \bra{u_{\bf k}^i}\hat z \ket{u_{\bf k}^j}$, with $ i =j$ cases have already been excluded in Eq.~(\ref{alphaxx}). Note that the simple form of position operator relies on Eq.~(\ref{blount}), which requires no band touching at the Fermi surface.

It may not be obvious at first glance what this result means physically, or how it can be connected to known formulas for the 3D response in terms of the orbital magnetic moment.  Hence we next give a diagrammatic explanation for the result, then apply it to slabs of increasing size to see how the 3D limit emerges quantitatively.

\section{Diagrammatic interpretation}\label{Diagrammatic}

\begin{figure}[!t]
\centering 
\includegraphics[width=1\columnwidth]{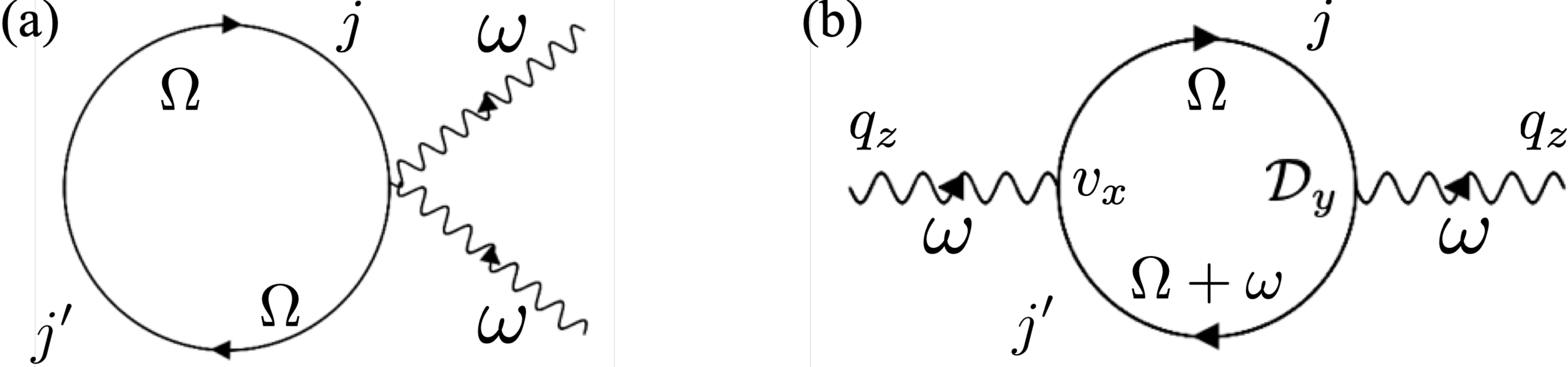}
\caption{\label{Diagrams}Two diagrams for Hamiltonian Eq.~(\ref{pert}) to the leading order. The solid line and wavy line, denotes, respectively, the electron and photon propagator~\cite{Dan2019}. (a) The dia-magnetic current. (b) The diagram for Eq.~(\ref{builddiagram}). The ${\mathcal D}_y$ stands for the vertex for perturbation as defined in Eq.~(\ref{Dy}) and $v_x$ stands for the vertex for the measurement.}
\end{figure}

In this section, similar to recent work on electric dipole responses~\cite{Dan2019,holder2019}, we would like to assign a diagrammatic interpretation for formula  Eq.~(\ref{alphaxx}). Note that the Hamiltonian can be written perturbatively as:
%\begin{widetext}
\begin{align}\label{pert}
	&H(\hbar {\bf k} + e \vec A) \nonumber\\
	&= \int [d {\bf k}] c^\dagger_{\bf k} H_0(\hbar {\bf k}) c_{\bf k} + \int[d {\bf k}] c^\dagger_{\bf k} \bigg{(}e \frac{\partial H}{\partial \hbar {\bf k}}\cdot{}{\bf A} \bigg{)}c_{\bf k}+ \cdots 
	\nonumber \\
	&= \int [d {\bf k}] c^\dagger_{\bf k} H_0(\hbar {\bf k}) c_{\bf k} + \int [d {\bf k}] c^\dagger_{\bf k}( e \hat {\bf v} \cdot{} {\bf A} )c_{\bf k} + \cdots.
\end{align}

%\end{widetext}
Here $c^\dagger_{\bf k}$ ($c_{\bf k}$) stands for the creation (annihilation) operator for a Bloch electron. (Since the speed of light $c$ is much larger than the Fermi velocity $\hat v_F$, we only ensure the energy conservation at each vertex~\cite{Dan2019}.) Here $[d{\bf k}] = dk_x dk_y/A_{\text{BZ}}$, with $A_{\text{BZ}}$ stands for the area of the 2d first Brillouin zone in which the integral is conducted. The first term is the unperturbed Hamiltonian, and the rest of the terms are perturbations from the external electromagnetic field. The amplitude of the diagram tells the response of one vertex (measurement) to the the other vertex (perturbation).  Let us imagine that we inject the light perpendicular to the $x$-$y$ plane (along $z$ direction). The magnetic field is in-plane. In Landau gauge $B_x = {\bm \nabla} \times {\bf A} $, where ${\bf A}= (0, zB,0)$. Thus the contribution is from $e \hat {\bf v} \cdot{}  {\bf A}$, and we can write this vertex in the canonical form of position operator as:
\begin{equation}\label{Dy}
	{\mathcal D}_y = e \hat {\bf v}\cdot{}  {\bf A} = \frac{eB}{2} ( \hat v_{{\bf k},y} \hat z  +  \hat z \hat v_{{\bf k},y}).
\end{equation} 
With the above we can figure out the Feynman diagram for a Bloch electron coupled to external electrical magnetic field, as shown in Fig.~\ref{Diagrams}. The Green's function or the propagator of the electron is defined as~\cite{Dan2019}:
\begin{equation}
	G (\omega)=(\omega -H({\bf k}))^{-1} = \sum_i \frac{\ket{u_{\bf k}^i}\bra{u_{\bf k}^i}}{\hbar \omega - E_{\bf k}^i}. 
\end{equation}
The first diagram is the diamagnetic current, same as in Eq.~(\ref{Diamagnetic}). The contribution from the second diagram reads:
\begin{equation}\label{builddiagram}
	\begin{aligned}
		&j_x^{\text{total}} = e \bigg{[} \sum_{j^\prime j } \int [d{\bf k}] \int d\Omega \langle{\hat v_{{\bf k}, x}}\rangle^{j^\prime j}_{\bf k} G_j (\Omega)\langle{{\mathcal D}_y}\rangle^{j j^\prime}_{\bf k} G_{j^\prime}(\Omega+\omega)\bigg{]}.
	\end{aligned}
\end{equation}
Note that~\cite{Dan2019}:
\begin{equation}
	I_2 = \int d\Omega G_j(\Omega) G_{j^\prime}(\Omega + \omega) = \frac{f^0(E_{\bf k}^{j^\prime}) - f^0(E_{\bf k}^j)}{\hbar(\omega + i/\tau) -(E_{\bf k}^{j^\prime}- E_{\bf k}^j)}.
\end{equation}
By using the same trick mentioned in the previous chapter, part of $I_2$  will cancel with the diamagnetic current, while the remaining part contributes to the GME current. In the low frequency limit, the part we are interested in for $I_2$ is $ \delta_{j^\prime j} \partial f^0(E_{\bf k}^j) /\partial {E_{\bf k}^j} [(i\omega \tau)/(1 - i\omega \tau)]$. We further have the current:
\begin{equation}
	j_x^{\text{total}} =  \frac{i\omega \tau e}{1 - i\omega \tau} \sum_j \int [d{\bf k}]  \frac{\partial f^0(E_{\bf k}^j)}{\partial E_{\bf k}^j}\langle \hat v_{{\bf k}, x} \rangle^{jj}_{\bf k} \langle {{\mathcal D}_y}\rangle^{jj}_{\bf k},
\end{equation}
from which we can subtract the $\alpha_{xx,2d}^{\text{GME}}$ as:
\begin{widetext}
\begin{equation}
	\alpha_{xx,2d}^{\text{GME}} = \frac{i\omega \tau }{1 - \omega \tau} \sum_{j} \int [d{\bf k}]  \frac{\partial f^0(E_{\bf k}^j)}{\partial E_{\bf k}^j} \langle \hat v_{{\bf k}, x} \rangle^{jj}_{\bf k} \sum_{j_1} \frac{e^2}{2} \big{[}\langle \hat v_{{\bf k}, y} \rangle^{jj_1}_{\bf k} \langle \hat z \rangle^{j_1 j}_{\bf k} + \langle \hat z \rangle^{jj_1}_{\bf k} \langle \hat v_{{\bf k}, y}\rangle^{j_1j}_{\bf k}\big{]},
\end{equation}	
\end{widetext}
which is equivalent to Eq.~(\ref{alphaxx}), aside from it contains $j = j_1$ term. Note that the terms within the brackets are already real. The result above corresponds to the total current, but the GME is related to its anti-symmetric part, so we need to drop the $j = j_1$ since it is invariant under $v_x \leftrightarrow v_y$. This is a straightforward way to understand the result from the Kubo formula without going through a rigorous calculation.

\section{Connection with 3d bulk results in thermodynamic limit}\label{Convergence}

Let us consider a stack of $N_z$ layers of the quasi-2d chiral structure along the $\hat e_z$ direction, as shown in Fig.~\ref{OpticalRotation}(b). In the thermodynamic limit, i.e., $N_z \rightarrow \infty$, we should have Eq.~(\ref{alphaxx}) converging to 3d bulk results~\cite{Ma2015,Zhong2016}:
\begin{subequations}\label{3dfull}
\begin{align}
	\alpha_{xx,3d}^{\text{GME}} &= \frac{i\omega \tau e}{(1 - i\omega \tau)} \sum_n \int [d{k}] \frac{\partial f^0(E_{ k}^n)}{\partial E_{k}^n} \hat {v}_{{k},x} \hat {\bf m}_{{k}n,x}, \label{3dbulk} \\
	\hat {\bf m}_{{k}n} &= \frac{e}{2\hbar } \text{Im} \bra{\partial_{ k} u_{ k}^n} \times (H_{ k} - E_{ k}^n) \ket{\partial_{ k} u_{ k}^n} ,\label{mbulkorbital}
\end{align}
\end{subequations}
with ${ k} = ({\bf k}, k_z) = (k_x,k_y,k_z)$, $[d k] = dk_xdk_ydk_z/V_{\text{BZ}}$, where $V_{\text{BZ}}$ stands for the volume of the 3d first Brillouin zone in which the integral is conducted. We write $\ket{u_k^n}$ for the 3d cell-periodic part of Bloch states, which are the eigenstates of the Bloch Hamiltonian $H_k \ket{u_k^n} = E_k^n \ket{u_k^n}$. By applying Eq.~(\ref{blount}) to Eq.~(\ref{mbulkorbital}) one can obtain the orbital magnetic moment written in terms of velocity operators:
\begin{equation}\label{3dmoment}
	\hat {\bf m}_{{ k}n,x} = e \hbar \sum_{m, m\neq n} \text{Im} \bigg{[} \frac{\bra{u_{ k}^n} \hat v_{k,y} \ket{u_{ k}^m} \bra{u_{ k}^m} \hat v_{k,z} \ket{u_{ k}^n}}{E_{k}^m - E_{k}^n}  \bigg{]}.
\end{equation}

Before going to realistic models, we provide a straightforward way to understand the connection between the 2d result Eq.~(\ref{alphaxx}) and the 3d result Eq.~(\ref{3dbulk}). Applying Eq.~(\ref{blount}) along $\hat e_z$ direction, we have: $\hbar  \langle \hat v_{k,y} \rangle_{k}^{mn}\equiv \bra{u_{ k}^m} \hbar \hat v_{k,y} \ket{u_{k}^n}_{\text{Cell}}  = \bra{u_{k}^m} \partial_{k_z} H_{ k} \ket{u_{k}^n}_{\text{Cell}} = i(E_{ k}^m - E_{k}^n) \bra{u_{ k}^m} \hat z \ket{u_{k}^n}_{\text{INV}} \equiv i(E_{ k}^m - E_{k}^n) \langle \hat z \rangle_{k}^{mn}$, with the subscript INV standing for 3d bulk material with infinite volume.
Note that such a replacement is only valid for a 3d bulk material where an infinite integral is conducted~\cite{Gu2013}. Substituting this to Eq.~(\ref{3dmoment}) we have:
\begin{equation}\label{mxreal}
	\hat {\bf m}_{{k}n,x} = \frac{e}{2}\sum_{m, m\neq n} \text{Re}\big{[}\langle \hat v_{{k},y} \rangle^{nm}_{k} \langle \hat z \rangle^{mn}_{ k} + \langle  \hat z \rangle^{nm}_{ k} \langle  \hat v_{{ k},y} \rangle^{mn}_{ k} \big{]}, 
\end{equation}
with Eq.~(\ref{mxreal}) back to the $\alpha_{xx,3d}^{\text{GME}}$ in Eq.~(\ref{3dbulk}).  Finally we arrive at the 3d bulk formula in terms of position operator:
\begin{widetext}
\begin{equation}\label{INV}
	\alpha_{xx,3d}^{\text{INV}} = \frac{i\omega \tau}{1 - i\omega \tau} \sum_n \int [d{k}] \frac{\partial f^0(E_{k}^n)}{\partial E_{k}^n} \langle \hat v_{{k},x} \rangle^{nn}_{k} \frac{e^2}{2}\sum_{m, m\neq n} \text{Re}\big{[}\langle \hat v_{{ k},y} \rangle^{nm}_{ k} \langle \hat z \rangle^{mn}_{ k} + \langle  \hat z \rangle^{nm}_{k} \langle  \hat v_{{k},y} \rangle^{mn}_{ k} \big{]},
\end{equation}
 \end{widetext}
 which looks similar to Eq.~(\ref{alphaxx}).

\begin{figure*}
\centering 
\includegraphics[width=500.0pt]{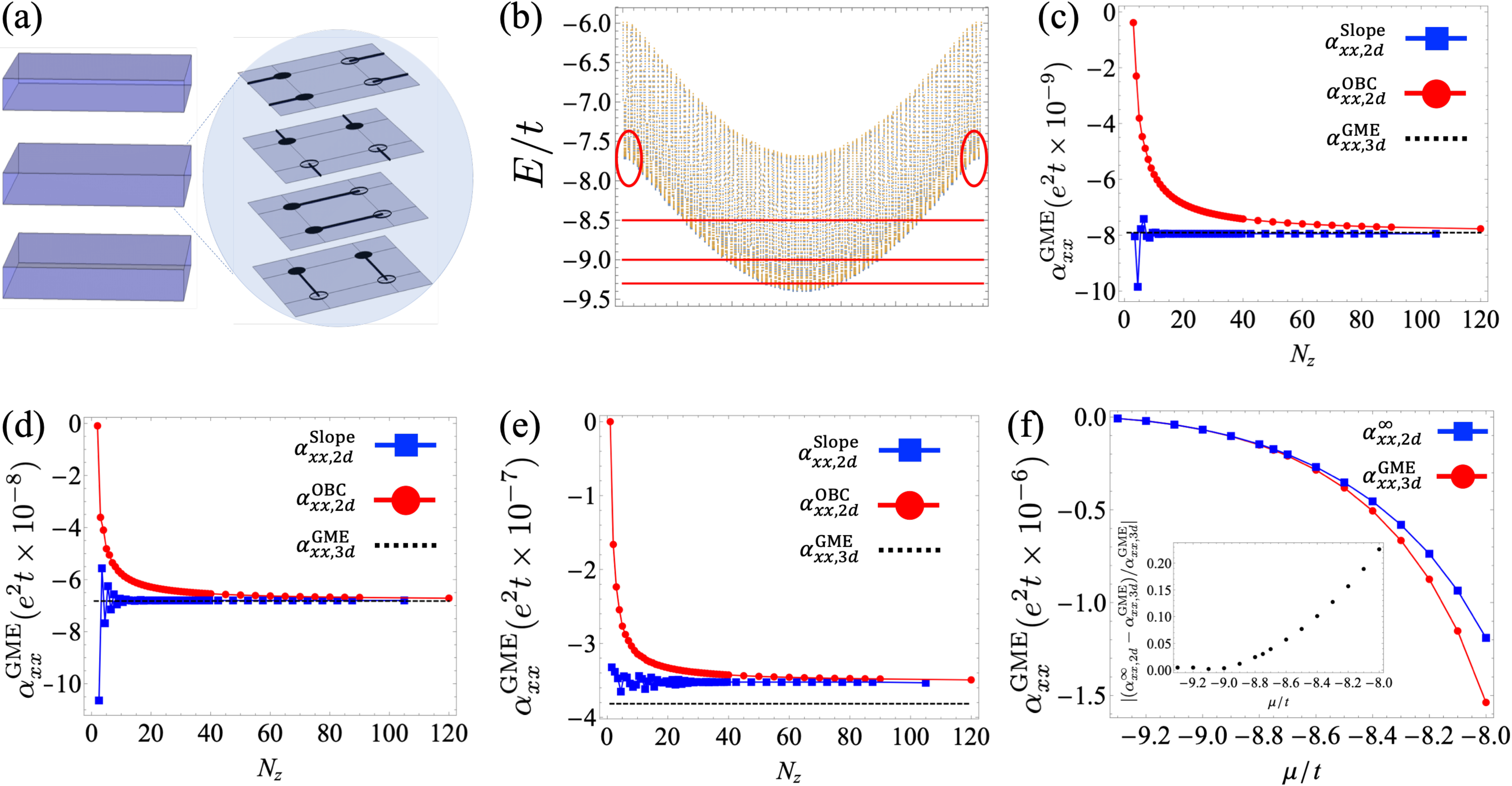}
\caption{\label{16bandsmodel}Sketch for the 16-band model and numerical results. (a) Left: the sketch for layer stacking; right: the configuration for each unit cell. Note that each layer itself has some structure along $\hat e_z$ direction. (b) The 2d band structure of a 15 layers slab for the lowest 7 subbands. Note that the 2d band structure should be understood as following: we assign each ${\bf k}$ point $(k_x,k_y)$ with a unique label, such that all ${\bf k}$ points form a one dimensional array. The $y$ axis stands for the energy, while the $x$ axis stands for the label for ${\bf k}$ points. The red lines in the figure from the bottom to the top stand for Fermi level in (c-e), respectively. The red circles highlight the band touching points. (c-e) The GME coefficient for the 16 bands model Eq.[\ref{16}], with $t = 1.0$, $t_z = 3.0t$, $\delta t = \epsilon= 0.4t$, and $k_BT= 0.03t$. The blue, red, and black line denotes, respectively, the slope/averaged value for $\alpha_{xx}$ based on Eq.[\ref{ave}], and 3d infinite volume results for Bloch electron based on Eq.[\ref{3dfull}]. The subfigures (c) $\mu = -9.3t$, the difference compared with 3d result is $|(\alpha_{xx,2d}^{\infty}-\alpha_{xx,3d}^{\text{GME}})/\alpha_{xx,3d}^{\text{GME}}| \approx $ 0.5\%,  (d) $\mu = -9.0t$, the difference compared with 3d result is $|(\alpha_{xx,2d}^{\infty}-\alpha_{xx,3d}^{\text{GME}})/\alpha_{xx,3d}^{\text{GME}}| \approx $ 0.4\%, (e) $\mu = -8.5t$, and the difference compared with the 3d result is $ |(\alpha_{xx,2d}^{\infty}-\alpha_{xx,3d}^{\text{GME}})/\alpha_{xx,3d}^{\text{GME}}| \approx $ 8.3\%. (f) The $\alpha_{xx,2d}^{\infty}$ and $\alpha_{xx,3d}^{\text{GME}}$ with respect to different Fermi levels. The relevant difference between 2d and 3d results is increased as more bands become close in energy, as shown in the subfigure.}
\end{figure*}

From Eq.~(\ref{alphaxx}) one can define two relevant variables:
\begin{subequations}\label{ave}
\begin{align}
&\alpha_{xx,2d}^{\text{OBC}}(N_z) = \frac{1}{N_z} \alpha_{xx,2d}^{\text{GME}}(N_z), \label{2dOBC} \\
&\alpha_{xx,2d}^{\text{Slope}}(N_z+\Delta/2) = [\alpha_{xx,2d}^{\text{GME}}(N_z + \Delta) -\alpha_{xx,2d}^{\text{GME}}(N_z)]/(\Delta). \label{2dSlope}
\end{align}
\end{subequations}
The $\alpha_{xx,2d}^{\text{OBC}}$ is just the layer averaged GME coefficient. When $\Delta = 1$, we have $\alpha_{xx,2d}^{\text{Slope}} \approx \partial \alpha_{xx,2d}^{\text{GME}}/\partial N_z$, which stands for the increasing of GME coefficient for an additional layer based on a $N_z$-layer slab. In large $N_z$ limit, $\alpha_{xx,2d}^{\text{OBC}}$ and $\alpha_{xx,2d}^{\text{Slope}}$ will converge to the same value by their definition, and we denote that as $\alpha_{xx,2d}^{\infty}$, i.e.:
\begin{equation}
   \alpha_{xx,2d}^\infty = \lim_{N_z \rightarrow \infty}\alpha_{xx,2d}^{\text{OBC}}(N_z)  = \lim_{N_z \rightarrow \infty} \alpha_{xx,2d}^{\text{Slope}}(N_z).
\end{equation}

In this section, we use a tight-binding model to verify that $\alpha_{xx,2d}^{\infty}$ converges to $\alpha_{xx,3d}^{\text{GME}}$ if the Fermi level is away from band touchings, which are inevitable in a time-reversal symmetric system.

Consider the following tight-binding Hamiltonian~\cite{Orenstein2013}:
\begin{equation}\label{16}
	{\mathcal H}_0 = -t\sum_{\langle ij \rangle} (c_i^\dagger c_j + c^\dagger_j c_i) - \delta t \sum_{b \in B} (c^\dagger_{b1}c_{b2} + c^\dagger_{b2}c_{b1}) + \epsilon \sum_{s \in S} c_s^\dagger c_s.
\end{equation}
Each unit cell contains 16 sites dispersed on 4 separated sheets along $\hat e_z$ direction, as shown in Fig.~\ref{16bandsmodel}(a). The subscripts $i$ and $j$ label sites of a nearest-neighbor bond with the nearest-neighbor hopping $t$. The nonzero onsite potential $\epsilon$ and thick bonds are added to make the model chiral and break the inversion symmetry. The thick bonds are labeled by $B$, the sites at two ends of a bond $b \in B$ are labeled by $b1$ and $b2$, and the set of solid-circle sites $S$. There is a screw axis parallel to $\hat e_z$ and passing through the upper left sites in Fig~\ref{16bandsmodel}(a).  Here, the distance between one site and its nearest neighbor is denoted as $a$.

In the quasi-2d formula Eq.~(\ref{alphaxx}), the position operator used in open boundary calculation reads: $\hat z = \{a, a, a, a, 2a, 2a, 2a, 2a,\cdots, 4N_za, 4N_za, 4N_za, 4N_za \}$. One can make this kind of unit cell periodic extensively in the $x$-$y$ plane, as in the blue layer shown in the left part of Fig.~\ref{16bandsmodel}(a), and stack $N_z$ identical layers which share the same screw axis along the $\hat e_z$ direction. The inter-layer coupling is just the  nearest-neighbor hopping $t$. For the open (free) boundary condition, the uppermost layer and lowest layer are decoupled, from which we can get $\alpha_{xx,2d}^{\text{OBC}}(N_z)$ and $\alpha_{xx,2d}^{\text{Slope}}(N_z)$ from Eq.[\ref{alphaxx}]. The results of $\alpha_{xx,2d}^{\text{OBC}}$, $\alpha_{xx,2d}^{\text{Slope}}$ and $\alpha_{xx,3d}^{\text{GME}}$ are presented in Fig.~\ref{16bandsmodel}(b-d) for different Fermi levels.

 One thing that we would like to point out is that Eq.~(\ref{blount}) (thus Eq.~(\ref{alphaxx})) does not apply if there is any degeneracy for 2d bands at certain $\tilde {\bf k}$. On the other hand, at these $\tilde {\bf k}$ there must be a value for Eq.~(\ref{simazb}) from using velocity operator $\hat v_z$, which may be different from the result if we directly use Eq.~(\ref{alphaxx}). One can see that as the Fermi level is tuned to the band bottom of lowest subband (away from the band touching points), the difference between the 2d results and 3d results is very small, in Fig.~\ref{16bandsmodel}(c-d,f).

\section{Conclusion}\label{Discussion}

In conclusion, based on standard perturbation theory, we derived a formula which evaluates the GME coefficient (i.e., optical rotation) for 2d thin chiral/twisted materials in the low-frequency limit. The formula is associated with the $\hat z$ position operator but extended states in the $x$ and $y$ directions, and can be easily applied in any 2d tight-binding model.  We further provided a Feynman diagrammatic interpretation for our formula, which helps to give it a straightforward physical meaning. Finally, we showed the convergence of the 2d formula in the thermodynamic limit to 3d bulk results.

The prediction of optical rotation based on this formula will be useful in current and future experiments, such as for determinining the chiralities of materials with different handedness and the size of the twist angle or, conversely, the rotation angle produced for a given twist.  Two possible extensions are to combine the results with tight-binding parametrizations produced by modern electronic-structure calculations and to generalize the results to the case of band touchings at the Fermi level and to Moir\'e systems without a unit cell.

\section*{\uppercase{Acknowledgements}}
We thank D. Parker, W. Berdanier, V. Bulchandani, A. Grushin, T. Cao, Z. Li and T. Xu for useful conversations. This work was primarily supported as part of the Center for Novel Pathways to Quantum Coherence in Materials, an Energy Frontier Research Center funded by the U.S. Department of Energy, Office of Science, Basic Energy Sciences.  
TM was supported by JST PRESTO (JPMJPR19L9) and JST CREST (JPMJCR19T3).

\bibliography{Paper_Yanqi_Layer_GME_v1}

\noindent

\end{document}